\documentclass[aps,prl,twocolumn,superscriptaddress]{revtex4}
\usepackage{amsmath,amssymb}
\usepackage{graphicx}
\usepackage{epstopdf}
\usepackage{bm}

\begin{document}



\title{Anisotropic spin fluctuations and superconductivity in ``115'' heavy fermion compounds: 
$^{59}$Co NMR study in PuCoGa$_5$}


\author{S.-H. Baek}
\email[]{sbaek.fu@gmail.com}
\thanks{current address: IFW-Dresden, PF 270116, 01171 Dresden, Germany}
\affiliation{Los Alamos National Laboratory, Los Alamos, New Mexico 87545, USA}
\author{H. Sakai}
\email[]{sakai.hironori@jaea.go.jp}
\affiliation{Los Alamos National Laboratory, Los Alamos, New Mexico 87545, USA}
\affiliation{Advanced Science Research Center, Japan Atomic Energy Agency,
Tokai, Ibaraki 319-1195, Japan}
\author{E. D. Bauer}
\affiliation{Los Alamos National Laboratory, Los Alamos, New Mexico 87545, USA}
\author{J.  N.  Mitchell}
\affiliation{Los Alamos National Laboratory, Los Alamos, New Mexico 87545, USA}
\author{J.  A.  Kennison}
\affiliation{Los Alamos National Laboratory, Los Alamos, New Mexico 87545, USA}
\author{F. Ronning}
\affiliation{Los Alamos National Laboratory, Los Alamos, New Mexico 87545, USA}
\author{J. D. Thompson}
\affiliation{Los Alamos National Laboratory, Los Alamos, New Mexico 87545, USA}

\date{\today}

\begin{abstract}
We report results of $^{59}$Co nuclear magnetic resonance measurements on a single crystal of 
superconducting PuCoGa$_5$ in its normal state.   
The nuclear spin-lattice relaxation rates and the Knight shifts as a function 
of temperature reveal an 
anisotropy of spin fluctuations with finite wave vector $q$. By comparison 
with the isostructural members, we  
conclude that antiferromagnetic XY-type anisotropy of spin fluctuations plays 
an important role in 
mediating superconductivity in these heavy fermion materials.    
\end{abstract}



\maketitle

The observation of 
unconventional superconductivity in the heavy fermion (HF) compounds (e.g., CePd$_2$Si$_2$ 
\cite{mathur98} and CeRhIn$_5$ \cite{hegger00}) in 
proximity to a magnetic instability initiated the now well accepted belief that spin 
fluctuations (SF) mediate Cooper pairing in these materials. 
Recently discovered transuranic HF compounds 
PuCoGa$_5$ \cite{sarrao02}, PuRhGa$_5$ \cite{wastin03}, and 
NpPd$_5$Al$_2$ \cite{aoki07} develop superconductivity at temperatures nearly an order of 
magnitude higher ($T_c=18.5$ K in PuCoGa$_5$) 
than in the previously known Ce-, U-, and Yb-based HF materials.  
Nuclear quadrupole resonance (NQR) studies \cite{curro05} confirm that 
superconductivity in PuCoGa$_5$ is mediated by spin  
fluctuations, also providing an important bridge linking the physics between 
HF and high $T_c$ cuprate superconductors.  More importantly the actinide 
based superconductors enable the possibility to investigate the microscopic 
factors which influence superconductivity within a single structural family of 115 HF 
superconductors.   

In the SF-mediated superconductors, the anisotropy of local SF appears to be 
relevant to the symmetry of superconducting pairs. In general, while the spin-triplet 
($p$-wave) superconductivity  
favors Ising-type coupling since only longitudinal fluctuations can induce an 
attractive force \cite{monthoux01}, the spin-singlet ($d$-wave) 
superconductivity prefers rather isotropic coupling since both longitudinal and 
transverse fluctuations can mediate Cooper pairing. In cuprates, the local 
SF is indeed isotropic in the normal state \cite{kastner98}. 
We show in this Letter, via the $^{59}$Co NMR, 
that the XY-type anisotropy of AFM SF scales  
with $T_c$ in the 115 HF superconductors, in striking contrast to 
the case of cuprates. Possible origins for this unexpected correlation are 
discussed.

NMR is an ideal local probe
since the spin-lattice relaxation rate ($T_1^{-1}$) is quite sensitive to 
these spin fluctuations.  
Generally, $T_1^{-1}$ is expressed \cite{moriya63} in terms of the dynamical 
susceptibility $\chi(\mathbf{q},\omega_{n})$ and hyperfine coupling  
$A$ whose components are perpendicular to the quantization axis: 
\begin{equation}
\label{eq:T1}
(T_1T)_\parallel^{-1} \propto 
\sum_\mathbf{q} [\gamma_n A_\perp(\mathbf{q})]^2\chi_\perp''(\mathbf{q},\omega_n)/\omega_n,
\end{equation}
where $\chi''$ is the imaginary part of $\chi(\mathbf{q},\omega_n)$,
$\omega_n$ is the nuclear Larmor frequency,
and the symbols $\parallel$ and $\perp$ denote the direction with respect 
to the quantization axis. 
The $\mathbf{q}$-dependent $A(\mathbf{q})$ can be approximated as 
$A(0)f(\mathbf{q})$, because the hyperfine coupling is local near the nucleus. In this relation,
 $A(0)$ is the hyperfine coupling constant and $f(\mathbf{q})$ is the hyperfine form 
factor determined by the geometrical configuration of nuclear sites.
Because the hyperfine coupling constant $A(0)$ is determined
from a linearity between the NMR shifts ($\mathcal{K}$) and the static 
susceptibility $\chi(0,0)\equiv\chi$ for each direction of the applied field $H$,
exact alignment of the sample with respect to $H$
is required. To 
prevent possible radioactive contamination during these experiments, the
single crystal of $^{239}$PuCoGa$_{5}$ must be encapsulated, making it very 
difficult to confirm the alignment of  
the sample after the encapsulation.  Here we take advantage of 
the quadrupole perturbed spectrum of $^{59}$Co ($I=7/2$) which is very 
sensitive to the angle between the 
applied field and the nuclear principal axis. 
For the axial symmetry, we expect seven spectral lines for $I=7/2$ which, in 
first order perturbation, should be 
equally separated by $\Delta\nu(\theta)=\nu_{Q}(3\cos^2\theta-1)/2$,
where $\theta$ is the angle between the 
principal $c$-axis of the electric field gradient (EFG) at the $^{59}$Co and 
the external field $H$ and $\nu_Q$ is the 
nuclear quadrupole frequency. By examining the $^{59}$Co spectra for 
 $H\parallel c$ and $H\perp c$ shown in 
Fig.~1 (a) and (b), misalignment of the sample for each 
direction, if any, is within $3^\circ$. 
We also determine the nuclear quadrupole frequency $\nu_{Q}=1.02$ 
MHz, which is comparable to $\nu_Q$ found in other 115 compounds 
\cite{sakai07,kambe07a}. 

\begin{figure}
\centering
\includegraphics[width=3in]{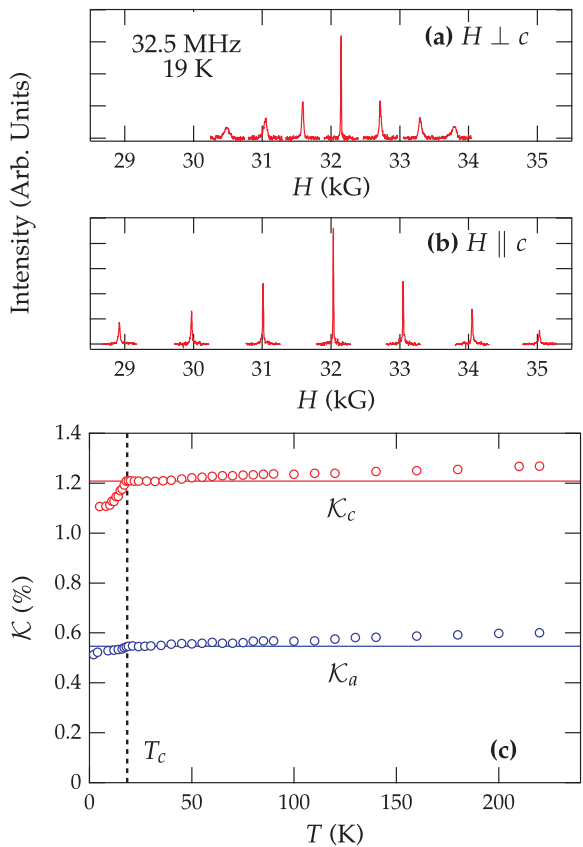}
\caption{\label{fig:spectra}
(a) and (b) $^{59}$Co NMR spectra at 19 K obtained by 
sweeping the external field $H$ at a fixed frequency 32.5 MHz.  (c) Knight shifts of the 
central transition for $H \parallel c$ and $H \perp c$. For 
$H\perp c$, a second order quadrupole correction was made, which is given by 
$\Delta\nu=(15/16)(\nu_Q^2/\nu_0)\sim 0.03$ MHz, or $\sim 0.09$ \%, where 
$\nu_0$ is the resonance frequency.}
\end{figure}

For measurements of $\mathcal{K}$,
the central transition ($\frac{1}{2}\leftrightarrow -\frac{1}{2}$)
is tracked as a function of temperature, shown in Fig.~1 (c).
Both $\mathcal{K}_c$ and 
$\mathcal{K}_a$ show similar temperature dependencies in the normal state: 
$\mathcal{K}_{a,c}$ decreases slightly with decreasing $T$, but becomes $T$-independent 
below $\sim 40$ K. At $T_{c}$ both shifts drop sharply, indicating spin-singlet pairing.
From the extrapolated zero-temperature values, $\mathcal{K} (T\rightarrow 0)$, 
we can estimate the orbital shift $\mathcal{K}_0$;
$\mathcal{K}_{0a}=0.5$ \% and $\mathcal{K}_{0c}=1.1$ \%.
The difference $\{{\mathcal K}-{\mathcal K}_{0}\}_{a,c}$ corresponds to the 
temperature-dependent spin part of $\mathcal{K}_{a,c}(T)$.
\nopagebreak[4] 
These $\mathcal{K}_{a,c}$--$T$ behaviors seem to be inconsistent with earlier results 
\cite{curro05}.
Although the origin of this discrepancy is not clear, 
recent polarized-neutron diffraction measurements on $^{242}$PuCoGa$_5$ \cite{hiess08} 
indicate a small, weakly temperature dependent static susceptibility, 
which suggests itinerancy of $5f$ electrons in PuCoGa$_{5}$.
Unlike the anisotropy found in $\mathcal{K}_{a,c}$, static susceptibility measurements 
on the same sample used in this work do not show anisotropy, which also is the case
with PuRhGa$_{5}$ and 
UCoGa$_{5}$ \cite{haga05,ikeda05}. We note, however, that reliable 
measurements of the uniform $\chi$ were complicated due to 
(i) encapsulation of the sample, (ii) Co impurities,
and (iii) radioactive damage from the decay process of Pu ($^{239}$Pu 
$\rightarrow$ $^{235}\mathrm{U}+\alpha$).
To check its order of magnitude, we roughly estimate 
$A_{a,c}=\mathcal{K}_{a,c}/\chi_{a,c}$ using  
the reported uniform $\chi$ \cite{hiess08}. This estimate gives 
$A_{a,c}$ in the range 5 to 10 kOe/$\mu_B$, which is close to
values found in UCoGa$_{5}$ \cite{kambe07a} and NpCoGa$_{5}$ \cite{sakai07}.

\begin{figure}
\centering
\includegraphics[width=3in]{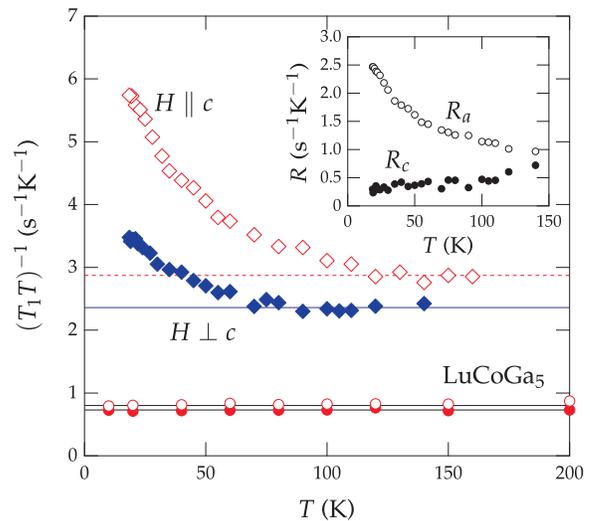}
\caption{\label{fig:invT1T}
Nuclear spin-lattice relaxation rate divided by $T$, $(T_1T)^{-1}$, as 
a function of $T$. For comparison, $^{59}$Co NMR of the nonmagnetic metal 
LuCoGa$_5$ is presented (filled circle: $H\perp c$ ; empty circle: 
$H\parallel c$). INSET: A plot of the in-plane component of fluctuations ($R_a$), 
which increases rapidly with  
decreasing $T$, and the out-of-plane component ($R_c$), which is almost 
independent of $T$.}
\end{figure}

The $T$-dependence of the nuclear spin-lattice relaxation rate divided 
by $T$, $(T_1T)^{-1}$, is plotted in Fig.~2 for $H\parallel c$ and 
$H\perp c$. 
Though both $(T_1T)^{-1}_\parallel$ and $(T_1T)^{-1}_\perp$ become $T$-independent
with a small anisotropy at high temperatures, 
both increase with decreasing $T$ and are 
accompanied by an increasing anisotropy $(T_1T)^{-1}_\parallel/(T_1T)^{-1}_\perp$
that reaches a maximum just above $T_c$.
In contrast, $^{59}(T_1T)^{-1}$ for LuCoGa$_{5}$ with its filled $f$ shell 
shows a very small and nearly isotropic $(T_1T)^{-1}$, as shown in Fig.~2.
Thus, the $T$-independent $(T_1T)^{-1}$ in 
PuCoGa$_5$ at high temperatures  
should originate from itinerancy of Pu's $5f$-electrons and not from conduction 
electrons. On the other hand,
the enhancement of $(T_1T)^{-1}$ below 100 K implies the partially localized 
nature of the $5f$ electrons. These observations may suggest evidence for a dual 
nature of $5f$ electrons in PuCoGa$_5$, which was previously implied from  
photoemission experiments \cite{joyce03}. It is noteworthy  
that, among the 115 HF 
superconductors, a $T$-independent $(T_1T)^{-1}$ at high temperatures has been 
observed \textit{only} in the  
Rh analog PuRhGa$_5$ \cite{walstedt07}, suggesting a unique feature of Pu-based materials.  
\begin{figure}
\centering
\includegraphics[width=3in]{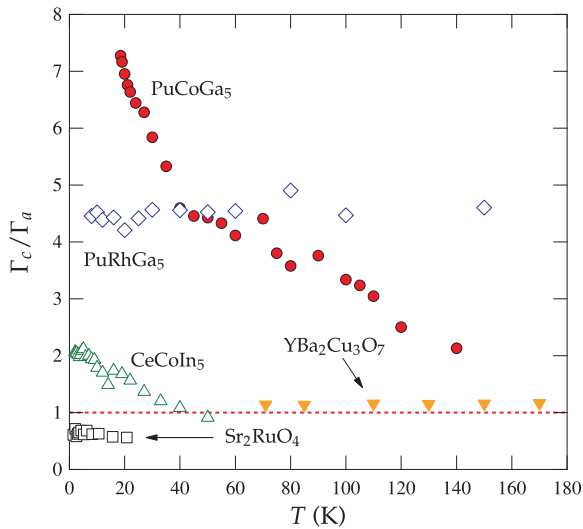}
\caption{\label{fig:gamma}
Ratio of spin fluctuation energy $\rho\equiv\Gamma_c/\Gamma_a$ as a function of 
temperature in the normal state.  Shown for comparison are results from $^{69}$Ga 
NMR in PuRhGa$_5$, $^{59}$Co NMR in CeCoIn$_5$ \cite{sakai10}, $^{101}$Ru NMR 
in Sr$_2$RuO$_4$ \cite{ishida01}, and $^{63}$Cu(2) NMR in YBa$_2$Cu$_3$O$_7$ 
\cite{walstedt89,pennington89}.}  
\end{figure}

Given $T_1^{-1}$ and $\mathcal{K}$, it is possible to estimate the magnetic 
nature of the spin fluctuations 
through the Korringa ratio defined as $R_\text{K} \equiv S/(T_1T)\mathcal{K}^2$, where 
$S= \mu_B^2/(\pi\hbar\gamma_n^2 k_B)$. In a simple metal or noninteracting 
Fermi gas, $R_\text{K}\sim 1$, but this ratio 
deviates from unity when electron-electron correlations are present  
\cite{moriya63,narath68}. For AFM fluctuations (i.e., magnetic fluctuation at finite $\mathbf{Q}$), 
$R_\text{K}$ becomes larger than 1, but it  
tends to be smaller than 1 when dominated by ferromagnetic fluctuations.
From $\mathcal{K}(T)$ and the 5$f$-derived contribution $(T_1T)^{-1}_{f}$ 
obtained by subtracting $(T_1T)^{-1}$ of LuCoGa$_{5}$, 
we find that $R_\text{K}$ ranges from 5 to 16, 
indicating the presence of strong AFM fluctuations in PuCoGa$_5$. 

To discuss in more detail the anisotropic nature of the AFM SF in PuCoGa$_{5}$, 
it is convenient to define new   
spin-lattice relaxation rates that \emph{probe} SF along 
the quantization axis.
In the tetragonal structure ($a=b\neq c$) of PuCoGa$_5$, these rates are defined by 
$R_\alpha \equiv [\gamma_n A(0)]^2\sum_\mathbf{q} 
\chi_\alpha''(\mathbf{q},\omega_n)/\omega_n$,
where $\alpha=a,c$. Here the form factor $f(\mathbf{q})=1$ is assumed for 
simplicity as it is irrelevant to our discussion \cite{footnote1_pucoga5}. 
Then, from Eq.~(1) $(T_1T)_{H \parallel c}^{-1}=2R_a$ 
and $(T_1T)_{H \perp c}^{-1} =R_a+R_c$.
As shown in the inset of Fig.~2, 
the in-plane component $R_{a}$, which is always larger than the out-of-plane $R_{c}$,
becomes prominent with decreasing $T$,
while $R_{c}$ slightly decreases.
In the case of AFM fluctuations,
we may take the main weight of $\chi''(\mathbf{q},\omega_n)$ around a finite $\mathbf{Q}$
as $\langle\chi''(\mathbf{q},\omega_n)\rangle$,
where $\langle\dots\rangle$ denotes the $q$ average.
In the limit of strong correlations,  
the approximation
$\chi''(\mathbf{Q},\omega_n)/\omega_n = 2\pi\chi^2(\mathbf{Q}) = 1/2\pi\Gamma^2(\mathbf{Q})$ holds
\cite{moriya95}.
Thus, the spin fluctuation energy becomes \cite{kambe07b}
\begin{equation}
\label{gamma}
       \Gamma_\alpha =\frac{\gamma_n A_\alpha(0)}{\sqrt{2\pi R_\alpha}},
\end{equation}
where $\Gamma_\alpha = \sqrt{\langle\Gamma_\alpha^2(q)\rangle}$.  
Using $A(0)\sim$ 5--10 kOe/$\mu_B$ estimated above, we find the average of 
$\Gamma_{a,c}$ to be 4--8 meV, which is much larger than 0.5--1 meV 
in CeCoIn$_{5}$ ($T_c=2.3$ K) \cite{sakai10} but lies in the range of the values found in many 
actinide 115 compounds \cite{kambe07b}. Inelastic neutron scattering measurements are 
necessary to confirm $\Gamma$ and $\mathbf{Q}$.  

\begin{figure}
\centering
\includegraphics[width=3in]{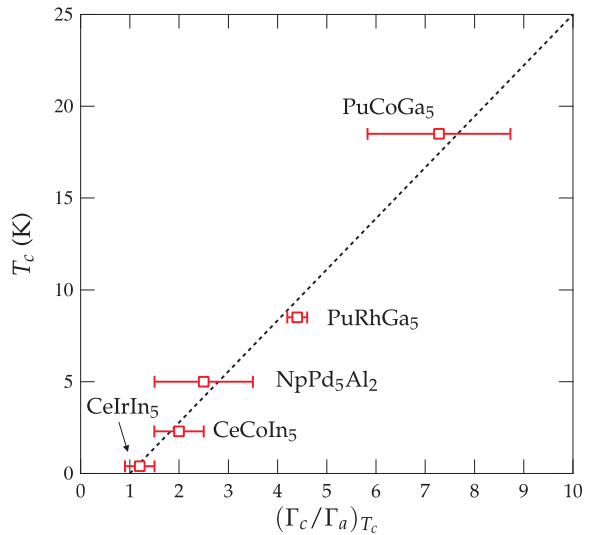}
\caption{\label{fig:scaling}
$T_c$ versus $\Gamma_c/\Gamma_a$ just above $T_c$ for 115 HF 
superconductors. Data for CeIrIn$_5$ and NpPd$_5$Al$_2$ are taken from 
refs.~\cite{kambe10} and \cite{chudo10}, respectively.   The dotted line is  
a guide to the eye, and the error bar for PuCoGa$_5$ is estimated assuming  
anisotropy of the static susceptibility is unity$\pm20$\%. } 
\end{figure}

Now we turn to the in-plane anisotropy of AFM SF in PuCoGa$_{5}$.
From Eq.~(\ref{gamma}) we define the anisotropy of $\Gamma$, 
\begin{equation}
\label{gamma_ratio}
\frac{\Gamma_c}{\Gamma_a} = \frac{A_c}{A_a}\sqrt{\frac{R_a}{R_c}} = 
\frac{\mathcal{K}_c(T)}{\mathcal{K}_a(T)} \sqrt{\frac{R_a}{R_c}}\frac{\chi_a}{\chi_c}.
\end{equation}
The ratio $\rho\equiv\Gamma_c/\Gamma_a$ is displayed in Fig. 3 as a function of 
$T$. We interpret this ratio as the anisotropy of SF which are 
peaked at $\mathbf{Q}$. Heisenberg systems such as the cuprates have $\rho \approx 1$ 
\cite{walstedt89,pennington89} while values less than 1 reflect Ising like anisotropy as is 
exemplified in the $p$-wave superconductor Sr$_2$RuO$_4$ \cite{ishida01}. In 
contrast, the $d$-wave superconducting 115 systems all have values of  
$\rho > 1$ which indicate XY like anisotropy. 
As noted above, $A_{a,c}$ cannot be determined accurately for PuCoGa$_5$; therefore, we 
express $A_{a,c}$ in terms of $\chi_{a,c}$ and $\mathcal{K}_{a,c}(T)$.  
$\chi(T)$ appears to be nearly isotropic, i.e.,
$\chi_a/\chi_c\sim 1$, and thus anisotropy in the spin fluctuation energy is 
dominated by $R_{a,c}$ and $\mathcal{K}_{a,c}(T)$. 
$\rho$ is a maximum just above $T_c=18.5$ K and shows an abrupt change at
$T^*\sim 60$ K, which corresponds to the hybridization gap observed in 
the photon-induced relaxation measurement \cite{talbayev10}.
As shown in Fig.~3, this behavior is somewhat similar to $\rho(T)$ observed in 
CeCoIn$_5$ \cite{sakai10} 
but different from that of PuRhGa$_5$. 
Clearly, $\rho$ just above $T_c$ for PuCoGa$_5$ is unprecedentedly large, 
much beyond the value in PuRhGa$_5$ that had been the largest $\rho$ among 115 compounds.

The primary result is presented in
Fig.~4, which shows the relationship between $T_c$ and $\rho$ 
just above $T_c$ for PuCoGa$_5$, PuRhGa$_5$ \cite{sakai05}, CeCoIn$_5$ \cite{sakai10}, 
CeIrIn$_5$ \cite{kambe10}, and NpPd$_5$Al$_2$ \cite{chudo10}.  
The error bar for  
$\rho$ of PuCoGa$_5$ is due to the estimate  $\chi_a/\chi_c=1\pm 0.2$,  
which should also include possible errors for $\mathcal{K}_c/\mathcal{K}_a$ in 
Eq.~(\ref{gamma_ratio}). The correlation between $T_c$ and $\rho$ shown in 
Fig.~4, in conjunction with the fact that $\rho\sim 1$ in  
nonsuperconducting 115 compounds \cite{kambe07a}, indicates that an increase 
of $T_c$ is associated with
more in-plane SF \cite{footnote2_pucoga5}. 
This result contradicts the expectation that Heisenberg systems should be more 
favorable for superconductivity due to the increased number of modes available 
to mediate pairing \cite{monthoux01}. 
A likely explanation is 
tied to the fact that spin-orbit coupling and crystal electric fields restrict 
the spin anisotropy in the 115 system. Consequently, the correlations found in 
Fig. 4 reflect the ability of the 115 compounds to optimize the spin 
anisotropy within the constraints of spin-orbit and crystal field 
interactions.  

We believe the most important parameter for setting the scale of $T_c$ is 
still the 
spin fluctuation energy scale $T_\text{SF}$, which explains why the superconducting 
transition temperature increases from Ce-based 115's to Pu-based 115's to 
pnictides to cuprates \cite{curro05}. In addition to $T_\text{SF}$, the 
reduced dimensionality of electronic correlations could also enhance $T_c$.  However, 
within 115 materials where  
$T_\text{SF}$, the correlation length ($\xi$) and its anisotropy 
($\xi_c/\xi_a$) are the same order of magnitude, the degree of XY anisotropy 
represented 
by $\Gamma_c/\Gamma_a$ is shown here to be a good parameter for determining $T_c$. It 
is surprising that both Ce-based 115s and Pu-based 115s lie on the same curve 
in Fig. 4. This may reflect the fact that due to spin-orbit coupling, spin 
anisotropy is naturally tied to the $c$-$f$ hybridization strength, which is a key 
parameter in setting the spin fluctuation energy scale. This gives a natural 
explanation for the observed temperature dependence of $\rho$ as well. 

In conclusion,
$^{59}$Co NMR measurements in the normal state of PuCoGa$_5$ have
uncovered the role of SF in promoting $d$-wave 
superconductivity in the isostructural 115  
HF compounds. Both the Knight 
shift $\mathcal{K}$ and the spin-lattice relaxation rate $T_1^{-1}$ show 
strongly anisotropic behavior. An analysis of the normal-state data finds   
an enhancement of SF at finite  
$\mathbf{Q}$ and strong in-plane (XY-type) anisotropy.
We suggest that the ratio $\Gamma_{c}/\Gamma_{a}$, a 
measure of the anisotropic spin fluctuations, 
is a characteristic quantity closely connected to the unconventional superconductivity 
in the 115 HF family.

We thank N. J. Curro, S. Kambe, S. E. Brown, H. Ikeda, and T. Takimoto for 
useful suggestions and discussions. 
H.S. acknowledges
the hospitality of Los Alamos National Laboratory.
Work at Los Alamos National Laboratory was performed
under the auspices of the U.-S. Department of Energy, Office of Basic Energy Sciences,
Division of Materials Sciences and Engineering and supported in part by the 
Los Alamos Laboratory Directed Research and Development program.

\bibliography{mybib}

\begin{thebibliography}{32}
\bibitem[{foo({\natexlab{b}})}]{footnote2_pucoga5}
\bibinfo{note}{$\rho$ values for non-superconducting compounds in their
  paramagnetic states are also consistent with the anisotropic nature of the
  ordered moments; e.g., for UPtGa$_5$ and NpCoGa$_5$ whose ordered moments
  point along the $c$ axis, $\rho$ is $\sim 0.68$ and $\sim 0.95$ respectively
  \cite{tokiwa02a,metoki05}, while $\rho\sim 1.7$ for NpFeGa$_5$ which exhibits
  in-plane AFM moments \cite{honda05}.}
\end{thebibliography}

\end{document}